# Power, Energy and Speed of Embedded and Server Multi-Cores applied to Distributed Simulation of Spiking Neural Networks: ARM in NVIDIA Tegra vs Intel Xeon quad-cores

Pier Stanislao Paolucci[1*], Roberto Ammendola[2], Andrea Biagioni[1], Ottorino Frezza[1], Francesca Lo Cicero[1], Alessandro Lonardo[1], Michele Martinelli[1], Elena Pastorelli[1], Francesco Simula[1], Piero Vicini[1]

[1]INFN Roma "Sapienza", Italy
[2]INFN Roma "Tor Vergata", Italy

[*]Corresponding author: Pier Stanislao Paolucci, E-mail pier.paolucci@roma1.infn.it

**ABSTRACT**

This short note regards a comparison of instantaneous power, total energy consumption, execution time and energetic cost per synaptic event of a spiking neural network simulator (DPSNN-STDP) distributed on MPI processes when executed either on an embedded platform (based on a dual-socket quad-core ARM platform) or a server platform (INTEL-based quad-core dual-socket platform). We also compare the measuer with those reported by leading custom and semi-custom designs: TrueNorth and SPiNNaker. In summary, we observed that: 1- we spent 2.2 micro-Joule per simulated synaptic event on the "embedded platform", approx. 4.4 times lower than what was spent by the "server platform"; 2- the instantaneous power consumption of the "embedded platform" was 14.4 times better than the "server" one; 3- the server platform is a factor 3.3 faster. The "embedded platform" is made of NVIDIA Jetson TK1 boards, interconnected by Ethernet, each mounting a Tegra K1 chip including a quad-core ARM Cortex-A15@2.3GHz. The "server platform" is based on nodes which are dual-socket, quad-core Intel Xeon CPUs (E5620@2.4GHz). The measures were obtained with the DPSNN-STDP simulator (Distributed Simulation of Polychronous Spiking Neural Network with synaptic Spike-Timing Dependent Plasticity) developed by INFN, that already proved its efficient scalability and execution speed-up on hundreds of similar "server" cores and MPI processes, applied to neural nets composed of several billions of synapses.

## 1. Introduction

Fast simulation of spiking neural network models plays a dual role: it contributes to the solution of a scientific grand-challenge – i.e. the comprehension of brain activity – and, by including it into embedded systems, it can enhance applications like autonomous navigation, surveillance and robotics. Therefore, this kind of simulations assumes a driving role, shaping the architecture of either specialized and general-purpose multi-core/many-core systems to come, standing at the crossroads between embedded and high performance computing. See, for example, [1], describing the TrueNorth low-power specialized hardware architecture dedicated to embedded applications, and [2] discussing the power consumption of the SpiNNaker hardware architecture, based on embedded multi-cores, dedicated to brain-simulation. See also [3], [4] as examples of approaches based on more standard High-Performance Computing platforms and general-purpose simulators. INFN APE lab developed a distributed neural network simulator and mini-application benchmark [5] in the framework of the EURETILE FP7 project [6][7][8]. Indeed, the Distributed Simulator of Spiking Neural Networks with synaptic Spike-Timing Dependent Plasticity mini-application has been developed with two main purposes in mind: as a quantitative benchmarking tool for the evaluation of requirements for future embedded and HPC systems, and as an optimized simulation tool addressing specific scientific problems in computational neuroscience. For example, in cooperation with ISS (Istituto Superiore di Sanità, Rome, Italy), in the framework of the CORTICONIC FP7 FET project [9], the distributed simulation technology is going to be used to accelerate the [10] simulator, to study slow waves in large scale cortical fields [11][12][13]; the next-to-start Horizon2020 EXANEST project will include DPSNN in the set of benchmarks used to specify and validate the requirements of future interconnects and storage systems. In this short note, we used this benchmark to compare "embedded multi-cores" and "server multi-cores" based on off-the-shelf components for what concerns: 1- total energy to solution (that required to complete a selected simulation task); 2- instantaneous power consumption during the simulation; and, 3- the ratio between total execution times on the platforms.

## 2. Methods

The first DPSNN technical report [5] described the strong and weak scaling behaviour of our DPSNN spiking net simulator: up to 256 MPI processes are distributed on the "server platform" QuonG [14] – based on a cluster of SuperMicro X8DTG-D 1U dual-socket servers – housing 128 computing cores residing on quad-core Intel Xeon CPUs (E5620@2.4GHz in 32nm CMOS technology with HyperThreading support, 2 threads per physical core). For this study, we used the InfiniBand switches and network cards, reserving a next work the optimization of the custom APEnet+ interconnect [15].

We used two NVIDIA Jetson TK1 boards, connected by an Ethernet 100Mb mini-switch, to emulate a dual-socket node. Each board mounts one NVIDIA Tegra K1 chip that includes a quad-core ARM Cortex-A15@2.3GHz in 28nm CMOS technology, a good representative of state-of-the-art "embedded multi-cores".

As reported in [5], the scaling of the DPSNN code has been verified on the "server platforms" on a range of configurations, from larger networks, composed from 6.6G synapses and 32.8M neurons, down to smaller configurations of 200K synapses and 1K neurons. Neural columns composed of 80% RS excitatory, 20% inhibitory FS Izhikevich neurons have been assembled in bidimensional grids, with synaptic connectivity decreasing with the distance between pairs of columns.

For this report, the power and energy consumption reported have been obtained simulating[1] 3s of activity of a network composed of 18M equivalent (internal + external) synapses: the network includes 10K neurons (leaky integrate-and-fire with Calcium mediated spike-frequency adaptation), each one projecting an average of 1195 internal synapses and receiving an "external" stimulus, corresponding to 594 equivalent external synapses/neuron. A Poissonian spike train targeted external synapses with average rate of 3Hz. Synaptic plasticity was disabled. In response, the neurons fired trains of spikes at a mean rate of 5.1Hz. On the "server platform" the minimum elapsed time is obtained assigning 16 MPI processes to each dual-socket, quad-core Intel Xeon server node. A similar "dual-socket" embedded node can be emulated using two Jetson TK1 boards, each one mounting a quad-core ARM Cortex-A15. In this last case, best execution time is obtained using 8 MPI processes, because of the missing support for HyperThreading.

We used an amperometric clamp to observe the currents flowing from the external 220V/50Hz power supply to the "embedded" and "server" platforms during simulation and measured the times required to complete identical simulation tasks. Note that we did not subtract any "base-line" power, e.g. power consumption after bootstrap, so the estimate is "pessimistic" and includes all system overhead. See the "Results" and "Discussion" sections.

## 3. Results

The simulation of 3s of activity of the neural network described in the "Methods" section required a time of 9.1s on the Intel Xeon E5620 dual-socket node and 30s on the emulated "embedded" dual-socket, based on the ARM Cortex-A15 in the NVIDIA Tegra. Observed currents were $I_s$ = 1.15A ("server") and $I_e$ = 80mA ("embedded"), with 5mA measure error. Therefore, the energy required to complete the same task on the "server" and "embedded" dual-socket nodes were $E_s$ = 2.3KJ and $E_e$ = 528J, while the observed instantaneous power consumptions were $P_s$ = 253W and $P_e$ = 17.6W.

The following section translates the measures into joule per synaptic event, compares the performances of the "embedded" and "server" multi-cores and gives pointers to measures performed by other groups.

---

[1] DPSNN svn rev 886 - 13 Feb. 2015

## 4. Discussion

The "server" dual-socket node is faster, spending 3.3 times less time than the "embedded" node. However, the "embedded" node consumes a total energy 4.4 times lower to complete the simulation task, with an instantaneous power consumption 14.4 times lower than the "server" node.

Table 1. Execution time, power and energy per synaptic event on ARM Cortex A-15 and INTEL Xeon E5620 quad-core based platforms measured using the DPSNN spiking neural net simulator.

|  | "SERVER PLATFORM" node | "EMBEDDED PLATFORM" node | EMBEDDED / SERVER COMPARISON |
|---|---|---|---|
|  | dual-socket, quad-core Intel Xeon CPU E5620@2.4GHz nodes, hosted by 1U SuperMicro X8DTG-D nodes | Emulated dual-socket, quad-core ARM Cortex A15@2.3GHz in Tegra, mounted on NVIDIA Jetson TK1 boards |  |
| Joule per synaptic event | 9.8μJ / syn. Event | 2.2μJ / syn. event | NOTE: power is measured directly at the 220V@50Hz plug, without excluding any "base-line" power (e.g. stand-by after bootstrap) |
| Total Energy to complete the task | 2.3KJ | 528J | "embedded" 4.4 times betted than "server" |
| Instantaneous over-all power consumption | 253W | 17.6W | "embedded" 14.4 times better than "server" |
| Time to complete the simulation task: | 9.1s | 30s | "server" 3.3 times faster than embedded |

The simulation produced a total of 235M synaptic events: the total energetic cost of simulation can be estimated in 2.2μJ/synaptic event on the "embedded" node. We measured the overall power consumption of the boards, directly on the 220V@50Hz plug, while the measures reported in [1] and [2] estimated the DC power consumption.

The energetic cost of the optimized Compass simulator of the TrueNorth ASIC-based platform, run on Intel Core i7 CPU 950@3.07GHz (45 nm process) with 4 cores and 8 threads, is 5.7μJ/synaptic event [1], but excludes a significant "base-line" power consumption, i.e. the stand-by power after bootstrap. Our measures show that if we excluded a similar "base-line" power consumption our power would be reduced by a factor 4 on the "server" platform, and by a factor 2 on the "embedded" platform. In addition, the DPSNN simulator executes single or double precision computations, its synapses are free to assume a range of floating point values and the number of projected synapses can grow to thousands per neuron; on the contrary, TrueNorth and its Compass simulator use single bits for synapses, reduced numerical precision in computations and project just a few hundreds of synapses per neuron. For completeness, we also mention that: 1- the cost of the SpiNNaker platform [2], based on its specialized ARM multi-core is 20nJ/synaptic event; 2- the cost of TrueNorth ASIC platform is 26pJ/synaptic event [2], trading-off versatility of general purpose architectures with higher efficiency of more specialized once. An expanded version of this note will report a more extensive discussion of errors of measure and more accurate comparison with other custom and general-purpose simulation systems. However, we guessed that this short and preliminary report could be anyhow of some interest.

## 5. Acknowledgements

The DPSNN simulator development has been funded by the EURETILE European project [8] and by the CORTICONIC project [9]. The measures reported in this short note have been performed in the framework of the INFN internal project COSA.